\documentclass{article}

\usepackage{vendor/iclr2027/iclr2027_conference,times}
\usepackage{amsmath,amssymb,booktabs,graphicx,microtype,needspace,placeins}
\usepackage{algorithm,algpseudocode}
\usepackage[hidelinks]{hyperref}
\usepackage{xurl}

\newcommand{\SYS}{\mbox{AgentReplay}}

\title{\SYS{}: Token-Wise Trace Replay\\
Is Essential for Fair Serving System\\
Performance Benchmarking}

\author{Zaifeng Pan$^{1}$\thanks{Correspondence to: Zaifeng Pan $<$zapan@ucsd.edu$>$.} \quad
Michael Wang$^{1}$ \quad
Chris Wu$^{1}$ \quad
Xinwei Qiang$^{1}$ \quad
Zhengding Hu$^{1}$ \quad
\\ \bf
Zhongkai Yu$^{1}$ \quad
Yufei Ding$^{1}$
\\
$^{1}$ University of California San Diego
}

\iclrfinalcopy

\begin{document}
\maketitle
\lhead{Preprint. Under review.}

\begin{abstract}
LLM-based agents execute multi-turn workflows with interleaved model inference and tool calls, making efficient serving increasingly important.
However, evaluating serving optimizations is challenging because identical tasks can produce different execution trajectories.
Changes in generated tokens can alter subsequent prompts, tool calls, and reasoning turns, making it difficult to distinguish system improvements from workload variation.
Greedy decoding does not guarantee identical outputs, while replaying only sequence lengths loses token information that affects prefix caching and mixture-of-experts (MoE) routing.
To address these problems, we propose \SYS{}, a configurable trace record-and-replay framework for agent serving.
\SYS{} records input/output tokens, expert selections, request dependencies, and tool durations.
During replay, it forces the recorded output tokens while performing normal autoregressive computation, with optional controls for MoE expert selection and tool delays.
This allows different systems to execute the same recorded workload while retaining their own batching, scheduling, and parallelization decisions.
We further separate trajectory generation from performance evaluation, enabling compatible smaller models to replay long-horizon traces collected with more capable models.
Our experiments show that token-wise replay in \SYS{} effectively eliminates workload variation that greedy decoding and length-wise replay cannot avoid, enabling fairer performance comparisons across serving configurations.
\end{abstract}

\section{Introduction}
\label{sec:introduction}

Large language model (LLM)-based agents solve complex tasks through repeated reasoning and tool use~\citep{yao2023react}.
Their multi-turn execution and growing contexts introduce substantial computation and memory costs.
To improve serving efficiency, existing systems optimize scheduling, batching, memory management, and decoding~\citep{vllm,sglang,leviathan2023fast}.
Fair evaluation requires both the baseline and optimized systems to execute the same workload.

However, using the same initial tasks does not ensure the same workload.
LLM outputs can vary across runs, and these outputs determine the agent's subsequent actions.
For example, a few different tokens can change a tool call or cause early termination, producing a different trajectory.
In such cases, a lower execution time may simply come from fewer tokens or reasoning turns rather than a more efficient system.
This makes it difficult to attribute performance gains to the optimization being evaluated.

Greedy decoding and length-wise replay only partially address this problem.
Setting the temperature to zero removes sampling randomness, but numerical differences across batch sizes and serving configurations can still change the highest-scoring token~\citep{yuan2026understanding,he2025nondeterminism,qiang2026dash}.
Length-wise replay fixes input and output lengths without preserving token content.
This can affect prefix caching: if a response differs from the recorded history in the next prompt, its KV cache may no longer be reusable~\citep{sglang,qin2025mooncake,liu2025lmcache}.
For mixture-of-experts (MoE) models, different tokens can also select different experts, changing load balance, memory access, and communication~\citep{yu2026patterns,yao2024exploiting,eplb}.
Therefore, matching sequence lengths alone does not ensure comparable execution work.

Besides LLM generation, tool execution introduces another source of variation.
A longer tool call delays the next LLM request and leaves its KV cache idle for longer.
This changes request concurrency and memory pressure, affecting performance comparisons even when the optimization targets only LLM serving.

To address these problems, we propose \SYS{}, a configurable trace record-and-replay framework for vLLM~\citep{vllm}.
The key idea is to replay recorded agent behavior while measuring how each serving configuration executes it.
Specifically, \emph{token-wise replay} supplies recorded input tokens and forces each output token while performing normal autoregressive computation.
\emph{MoE router replay} restores logical expert selections at each layer and token position, while \emph{tool-duration replay} reproduces the delay of each tool call.
The replay driver follows recorded request dependencies, allowing batching, scheduling, and completion times to change with the serving system.
Users can enable these controls independently.

We further use token-wise replay to support long-horizon evaluation with smaller models.
Small models reduce resource requirements but may fail before reaching long contexts or many interaction turns.
\SYS{} allows a compatible smaller model to execute trajectories collected with a more capable model.
The target tokenizer defines a fixed token workload that retains the recorded content and turn structure.
This supports lower-cost performance evaluation without requiring the smaller model to solve the original tasks.

Our experiments show that token-wise replay in \SYS{} effectively eliminates workload variation that greedy decoding and length-wise replay cannot avoid.
By preserving the recorded token sequences rather than only the initial tasks or sequence lengths, \SYS{} enables fairer comparisons of serving performance under the same agent workload.
This paper makes three contributions:
\begin{itemize}
    \item We identify how output and tool-time variation affects agent serving benchmarks and motivate token-wise replay for fair comparisons.
    \item We propose \SYS{}, supporting configurable token-wise, MoE router, and tool-duration replay while retaining model computation.
    \item We enable long-horizon serving evaluation with smaller models by separating trajectory generation from performance measurement.
\end{itemize}

\section{Background}
\label{sec:background}

\begin{figure}[t]
    \centering
    \includegraphics[width=0.94\linewidth]{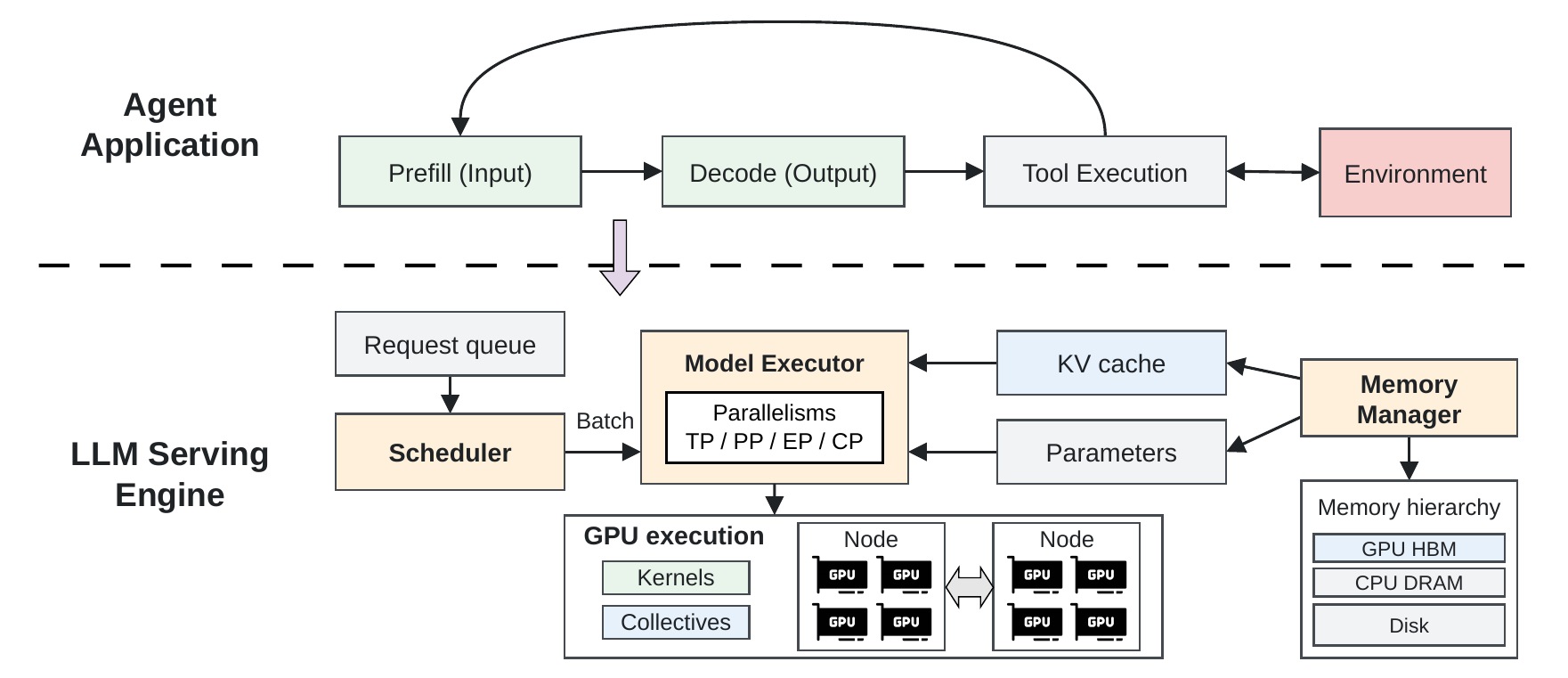}
    \caption{Agent execution and its LLM serving backend. The agent application repeatedly alternates between model inference and external tool or environment interactions. The serving engine queues and batches the resulting requests, executes the model under the configured parallelism, and manages parameters and KV caches across the memory hierarchy.}
    \label{fig:agent-serving}
\end{figure}

\paragraph{Agent workloads.}
LLM agents solve long-horizon tasks through repeated interactions with users, tools, and external environments~\citep{yao2023react,shinn2023reflexion,yang2024swe,wang2025openhands}.
As shown in the upper half of Figure~\ref{fig:agent-serving}, each agent turn submits the accumulated context to an LLM, which prefills the input and decodes an output token by token.
The output may answer the user, invoke a tool, or determine the agent's next action.
After a tool or the environment returns an observation, the agent appends it to the interaction history and starts another turn.
Requests from one agent are therefore stateful and intermittent: later turns depend on earlier model outputs and tool observations, while variable external work separates consecutive LLM invocations.

An execution graph captures dependencies among LLM and tool invocations, including concurrent branches.
Together with token sequences, tool delays, and task arrivals, it defines the workload observed by the serving engine.

\paragraph{LLM serving.}
The lower half of Figure~\ref{fig:agent-serving} shows the main components of a modern LLM serving engine.
Incoming requests first enter a queue, and the scheduler forms iteration-level batches to share accelerator capacity across prefill and decode~\citep{yu2022orca}.
The model executor runs GPU kernels and collectives under tensor, pipeline, expert, or context parallelism~\citep{shoeybi2019megatron,liu2024ringattention,jacobs2023deepspeed,fedus2022switch}, while the memory manager places model parameters and KV caches across GPU memory, CPU memory, and secondary storage.
Serving performance depends on decisions at each layer.
Prior work improves batching and request scheduling~\citep{yu2022orca,agrawal2024taming,zheng2024batchllm}, kernels and operator execution~\citep{ye2025flashinfer,dong2024flex,pan2025fasttree,dao2022flashattention,sanovar2025leanattention}, parallel and disaggregated execution~\citep{zhong2024distserve,patel2024splitwise,zhu2025megascale,liu2025deepseek}, and KV-cache management~\citep{qin2025mooncake,xie2026strata,liu2025lmcache,gao2024cost}.
Agent-oriented systems further exploit tool timing or workflow structure to schedule requests and manage cached state~\citep{abhyankar2024infercept,li2025continuum,pan2025kvflow,pan2026scalesim,chen2026concur,kang2026thunderagent,pan2026smoothagent}.
These optimizations must be evaluated under a controlled workload so that changes in agent behavior do not obscure system effects.

\paragraph{Content affects serving work.}
Sequence lengths determine tensor dimensions, but token contents also affect serving performance.
For example, prefix caching reuses KV tensors when requests share the same token prefix~\citep{vllm,sglang}.
Suppose the first turn generates a response $c$ that is included in the next prompt.
If length-wise replay generates $c'$ instead, the system may need to recompute the KV cache for $c$, even when $|c|=|c'|$.
In MoE models, the router selects a subset of experts for each token~\citep{yang2025qwen3}.
Different selections change the batch size of each expert and, with expert parallelism, the devices that exchange activations.
To compare expert placement or communication optimizations, we therefore distinguish which logical experts are selected from where and how they execute.

\paragraph{Timing affects subsequent work.}
Tool calls create gaps between dependent LLM invocations.
Many tools execute as CPU processes, perform file or network I/O, or invoke remote services.
OS scheduling, resource contention, and remote-service load make tool durations variable.
Real coding-agent traces accordingly exhibit tool-dependent, heavy-tailed latencies~\citep{zhu2026tracelab}.
During each gap, other requests continue executing, and the serving system may keep, offload, or evict the waiting agent's KV cache~\citep{abhyankar2024infercept,li2025continuum}.
A change in tool duration therefore affects both the arrival time of the next request and the memory pressure while the agent waits.
Tool timing matters even when the benchmark evaluates only the LLM serving system.

\paragraph{Benchmarking objective.}
A serving benchmark should expose each compared configuration to the same workload properties that affect the optimization under study, while leaving the serving mechanisms themselves free to change.
Given trace $\tau$, we compare its execution cost $C(S_1,\tau)$ and $C(S_2,\tau)$ under serving configurations $S_1$ and $S_2$.
The trace fixes the enabled workload properties while the serving engine controls batching, scheduling, cache placement, kernels, communication, and physical expert placement.
Repeating the same trace measures residual runtime variation without allowing a changed agent trajectory to redefine the work.
Live-agent evaluation remains necessary to assess task success and behavioral effects.
Length-wise replay is sufficient only when the mechanism under study is insensitive to token content and other omitted workload properties.

\section{\SYS{}}
\label{sec:method}

\begin{figure}[htb]
    \centering
    \includegraphics[width=0.94\linewidth]{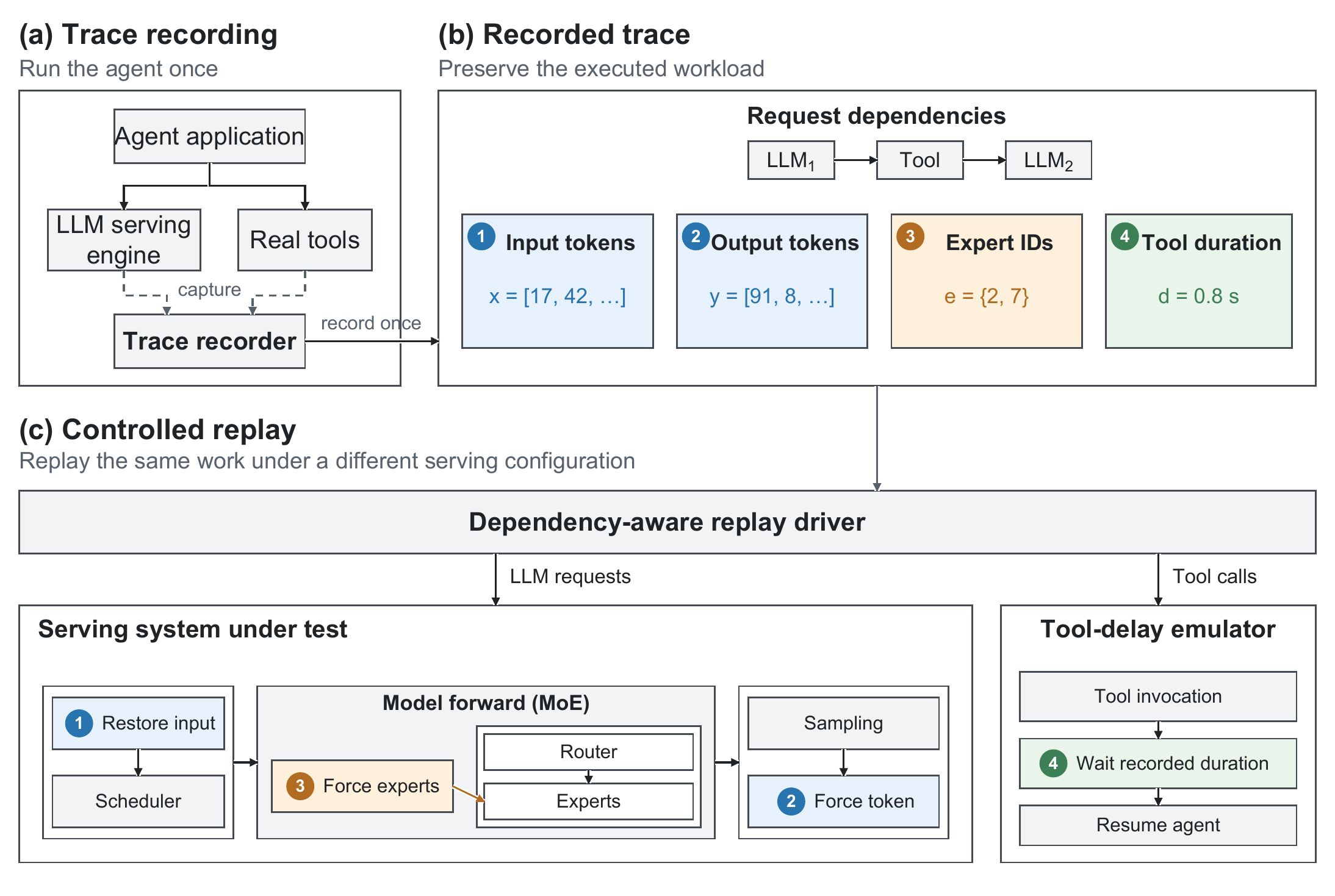}
    \caption{Overview of \SYS{}. (a) The recorder captures LLM and tool invocations during agent execution. (b) The trace stores request dependencies, input/output tokens, expert selections, and tool durations. (c) The replay driver follows these dependencies and submits requests to the serving system and tool-delay emulator. Users can select replay controls independently, while model computation, scheduling, and expert placement remain configurable. Token IDs and durations in the diagram are examples.}
    \label{fig:overview}
\end{figure}

Figure~\ref{fig:overview} shows the design of \SYS{}.
The recorder first collects a real agent execution, including LLM requests and tool calls.
The replay driver then uses this trace to evaluate different serving configurations.
Three replay controls preserve token sequences, logical expert selections, and tool delays, respectively.
Our evaluation uses vLLM~\citep{vllm}.

\subsection{Trace Recording}
\label{sec:trace}

We represent a trace as $\tau=(G,X,Y,E,D)$.
$G$ records invocation dependencies, $X$ and $Y$ store input and output tokens, $E$ stores optional logical expert selections, and $D$ stores tool durations.
The experiment separately specifies an arrival policy $A$, using either task release times or a fixed number of concurrent tasks.
To capture dependencies correctly, the recorder obtains them from the application or orchestration runtime rather than inferring them from execution timestamps.
The graph and recorded inputs also capture context edits, retries, branch decisions, and tool results.

Each invocation has a stable identifier, with metadata describing the model, tokenizer, prompt serialization, and sampling settings.
Inputs are stored as token IDs to avoid changes introduced by decoding and re-tokenizing text.
The recorder also retains termination reasons and, when needed for validation or application replay, tool names, arguments, and results.
During replay, the driver executes the recorded graph directly instead of asking the agent to choose its actions again.
As a result, tool-delay emulation does not repeat the external side effects of trace collection.

Replay fixes reuse opportunities and logical work, not the resulting cache hits, expert placement, or communication costs.
Those remain effects of the serving configuration under evaluation.

\subsection{Multi-Level Trace Replay}
\label{sec:replay}

\paragraph{Token-wise replay.}
\label{sec:token-replay}

For invocation $r$, the driver supplies input $X_r$ and the recorded output sequence $Y_r=(y_1,\ldots,y_m)$.
The model performs prefill to produce the first output token and then decodes autoregressively.
At output position $j$, the replay hook commits $y_j$ instead of the sampled token.
The hook replaces the token before updating the next-step input, request history, and token-dependent state.
Changing only the returned text would not reproduce the model workload.
Replay also follows the recorded stopping position, overriding conflicting EOS and stop rules while respecting the target's context limit.
Unsupported histories or decoding constraints are rejected rather than silently truncated.

Our design retains the forward pass, language-model head, and native sampling computation before replacing the selected token.
It therefore measures autoregressive execution rather than returning stored responses or processing all output tokens in one prefill pass.
Invocation IDs and output positions locate recorded tokens after batch reordering.

\paragraph{MoE router replay.}
\label{sec:router-replay}

Token-wise replay fixes the token sequence and recorded agent trajectory.
However, when benchmarking MoE optimizations, even small differences in expert selections can affect the comparison.
Identical tokens do not guarantee identical routing decisions, as numerical differences across execution configurations can change router scores~\citep{ma2025r3}.
\SYS{} therefore provides optional router replay for comparisons that require the same expert workload.
For top-$k$ routing, router replay fixes the logical expert IDs $E_{r,p,\ell}$ at token position $p$ and layer $\ell$ of invocation $r$.
The design retains router-score and native selection computation, then substitutes the recorded IDs before logical-to-physical expert mapping and dispatch.
The experts execute normally, while their placement and communication remain system decisions.
Fixing expert IDs does not require identical router scores or hidden states.

\paragraph{Tool-duration replay.}
\label{sec:tool-replay}

Tool-time variation changes subsequent arrivals and batch composition even when token sequences are fixed.
For the recorded graph, the replay driver releases each request after its required predecessors complete.
For a tool invoked at time $t$ with recorded invocation-to-result duration $d$, the emulator makes its result available after a nonblocking wait until $t+d$, subject to timer and scheduling error.
Independent tools wait asynchronously, while a join waits for all required predecessors.
Faster LLM execution can therefore release later requests earlier, rather than preserving their original absolute timestamps.

Many serving optimizations target GPU execution, while asynchronous scheduling overlaps serving-side CPU work with GPU computation~\citep{sglang2024overlap}.
When tool execution and its resource contention are outside the optimization scope, duration replay removes tool-time variation while retaining the waits and dependencies seen by the serving engine.
The serving engine still executes its own CPU scheduling and GPU computation normally.
This mode does not reproduce tool CPU work, I/O contention, or load-dependent service latency.
For systems that jointly optimize retrieval or other application stages with LLM generation~\citep{hu2025hedrarag,jin2025ragcache,tan2025towards,hu2026pancake}, users should disable tool-duration replay and execute the tools to measure these interactions.

\subsection{Long-Horizon Evaluation with Smaller Models}
\label{sec:small-model}

Resource constraints often motivate evaluation with smaller models, but their live executions may terminate early or enter repetitive tool loops.
These behaviors distort the context growth and reuse patterns of the intended long-horizon workload.
To address this problem, \SYS{} separates trajectory generation from performance evaluation.
A capable model first generates long trajectories, recording their prompts, responses, and tool dependencies.
A compatible smaller model then executes these trajectories through normal forward passes.
We compare serving configurations on this smaller model using the same traces, exposing context growth and token reuse that it may not reach through live generation.

Compatible tokenizers, special tokens, and prompt serialization allow direct token replay.
Otherwise, \SYS{} retokenizes the recorded messages and tool observations for the target, fixing that sequence across target-system comparisons.
This preserves content and dependencies rather than source token IDs or counts, subject to the target's context limit.
Expert IDs are architecture-specific, so router replay requires target-side route recording.

\section{Experiments}
\label{sec:experiments}

\paragraph{Setup and metrics.}
We evaluate Qwen3.6-35B-A3B on Astropy and SymPy agent tasks using vLLM.
Sections~\ref{sec:exp-greedy} and~\ref{sec:exp-cross-model} measure complete multi-turn runs, while Section~\ref{sec:exp-moe} measures batch trajectories.
For workload quantity $w$, we aggregate all requests in run $i$ into $w_i$ and report $\mathrm{CV}(w)=s(w_1,\ldots,w_n)/\bar{w}$, where $s$ is the sample standard deviation.
Thus, turns count LLM requests, and cumulative input tokens include context resubmitted across turns.
Model time sums request durations from first scheduling to the final token, and cumulative time adds tool durations.
Each complete run is one statistical sample, rather than treating correlated requests as independent observations.
Time summaries use mean $\pm$ standard deviation.

\subsection{Does Greedy Decoding Preserve the Workload?}
\label{sec:exp-greedy}

\begin{figure*}[t]
    \centering
    \includegraphics[width=\textwidth]{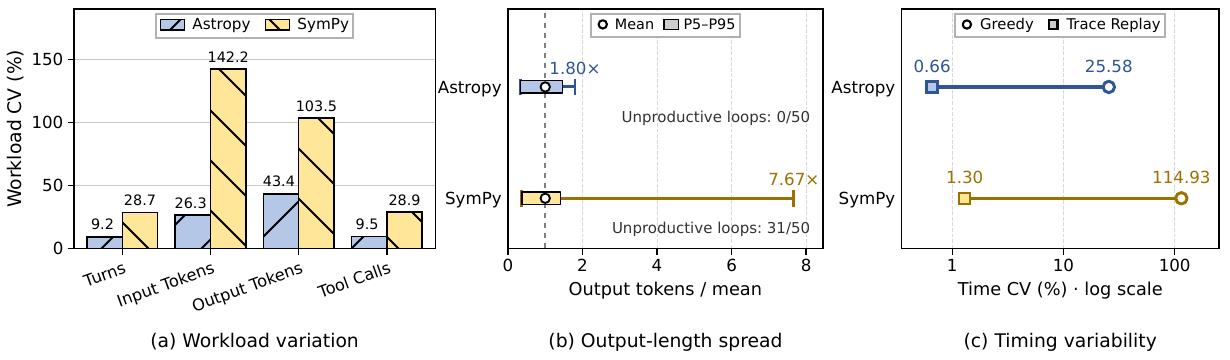}
    \caption{Greedy decoding changes the realized workload. (a) Run-level CVs for turns, input/output tokens, and tool calls. (b) Output tokens relative to the task mean; annotations report runs trapped in unproductive tool-calling or generation loops. Circles mark the mean, boxes span P5--P95, and whiskers show extrema. (c) CV of cumulative model-plus-tool time under greedy execution and trace replay (log scale). Each condition contains 50 complete runs.}
    \label{fig:greedy}
\end{figure*}

\paragraph{Greedy decoding changes the realized workload.}
We repeat each task 50 times with unchanged greedy-decoding settings.
Figure~\ref{fig:greedy}(a) shows substantial variation despite fixing the initial task and decoding configuration.
Astropy has CVs of 9.24\% in turns, 26.33\% in cumulative input tokens, and 43.38\% in output tokens.
SymPy varies more, with corresponding CVs of 28.66\%, 142.21\%, and 103.49\%.
Tool-call CVs of 9.51\% and 28.91\% further reflect changing control flow.

\paragraph{Divergence produces long-tailed episodes.}
In Figure~\ref{fig:greedy}(b), the largest Astropy and SymPy runs produce 1.80$\times$ and 7.67$\times$ their task's mean output tokens.
Among the 50 SymPy runs, 31 become trapped in unproductive cycles of repeated tool calls and text generation, compared with none of the Astropy runs.
Together with the long output tail, these cases show that greedy execution can repeatedly generate work without making progress toward task completion.

\paragraph{Trace replay stabilizes repeated measurement.}
Token-and-tool trace replay reduces cumulative-time CV from 25.58\% to 0.66\% on Astropy and from 114.93\% to 1.30\% on SymPy (Figure~\ref{fig:greedy}(c)).
The reductions, computed before rounding, are 38.9$\times$ and 88.7$\times$.
They measure repeatability rather than speedup because greedy runs and replay may contain different realized work.
Fixed tool durations also contribute to mean cumulative time without adding variance.

\subsection{Does Length-Wise Replay Preserve MoE Performance?}
\label{sec:exp-moe}

\begin{figure*}[t]
    \centering
    \includegraphics[width=\textwidth]{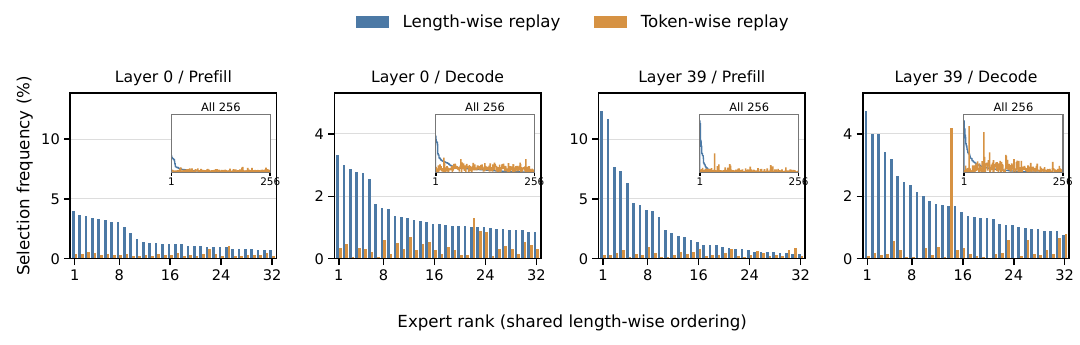}
    \caption{Expert selection under length-wise and token-wise replay. Panels show prefill and decode for layers 0 and 39. Experts are sorted by length-wise frequency within each layer and phase, with the same ordering for both modes. Main panels show the top 32 experts and insets show all 256. Frequencies sum to 100\% over all experts in each layer. Vertical scales match between layers for each phase.}
    \label{fig:expert-distribution}
\end{figure*}

\paragraph{Matching lengths changes the expert workload.}
Length-wise replay preserves input and output lengths but substitutes random tokens, whereas token-wise replay executes the recorded tokens.
Random tokens can concentrate routing on a subset of experts~\citep{ma2026dodoco,abramovich2026speedbench}.
Figure~\ref{fig:expert-distribution} compares layers 0 and 39 using a shared length-wise frequency ordering, so paired bars always represent the same expert.
Many experts favored by length-wise replay receive little traffic under token-wise replay.
Across all 40 layers, the top 32 length-wise experts receive 77.7\% of prefill assignments and 56.8\% of decode assignments on average.
The same experts receive only 13.8\% and 17.1\% under token-wise replay.
Even its own top 32 experts account for only 36.9\% and 43.9\% of token-wise assignments.
Thus, random tokens change both concentration and hotspot identities.

We examine the resulting costs across batch sizes 1--32 with expert parallelism (EP), tensor parallelism (TP), and expert offloading.
All times cover a complete batch trajectory, with prefill and decode reported separately.
Kernel totals include only \texttt{fused\_moe\_kernel} across 40 layers and exclude routing, activation, communication, and warmup.
Decode contains 188 steps.
Full-phase timings are measured separately without profiling.
Each ratio uses token-wise replay at the same batch size, phase, and configuration as its reference.

\begin{figure*}[t]
    \centering
    \includegraphics[width=\textwidth]{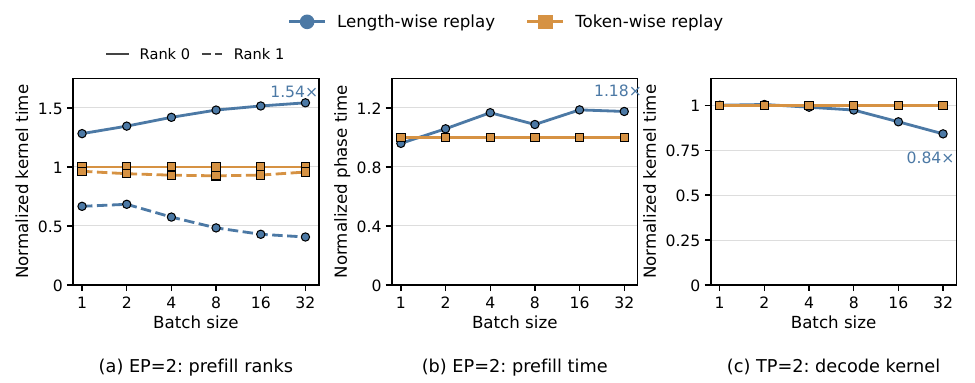}
    \caption{MoE execution times over a complete batch trajectory. (a) EP=2 prefill kernel totals per rank, normalized by the maximum token-wise rank total. Solid and dashed lines denote ranks 0 and 1. (b) Full prefill phase time with EP=2. (c) Maximum-rank decode kernel time with TP=2 and EP disabled. Token-wise replay is the reference in all panels.}
    \label{fig:moe}
\end{figure*}

\paragraph{Expert concentration creates a prefill straggler under EP.}
We enable EP=2 with TP=2 and fix a placement that co-locates length-wise hot experts on rank~0.
Figure~\ref{fig:moe}(a) normalizes both ranks by the maximum token-wise rank total.
The length-wise prefill rank-time ratio increases from 1.92$\times$ to 3.80$\times$, while token-wise replay remains within 1.04--1.08$\times$.
At batch~32, the sum across ranks differs by less than 0.4\%, but the larger length-wise rank total is 1.54$\times$ the token-wise reference.
The full prefill phase takes 1.18$\times$ as long (Figure~\ref{fig:moe}(b)).
This result exposes the interaction between token content and expert placement.

\paragraph{A smaller expert working set underestimates TP decode cost.}
With EP disabled and TP=2, we compare the maximum of the two ranks' cumulative decode kernel times (Figure~\ref{fig:moe}(c)).
The two modes are nearly identical at batches 1--4, but length-wise time falls to 0.91$\times$ and 0.84$\times$ the token-wise reference at batches 16 and 32.
Concentrating tokens on fewer experts can reduce distinct weight accesses and increase reuse in memory-bound decode GEMMs.
At batch~1, both workloads select eight experts per token, leaving little scope for within-batch reuse.
Counters from the offload experiment support the working-set explanation: at batch~32, length-wise replay touches 94.53 distinct experts per layer and decode step, compared with 117.27 for token-wise replay.
These counters characterize access patterns but do not directly establish the TP kernels' bandwidth utilization.

\begin{figure*}[t]
    \centering
    \includegraphics[width=\textwidth]{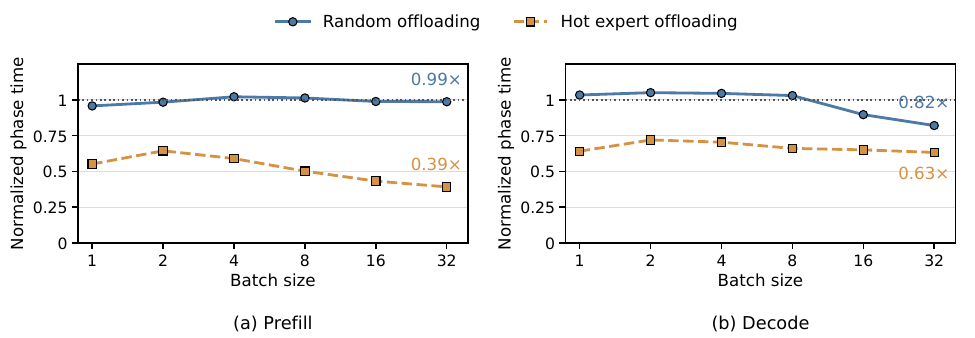}
    \caption{Offloaded execution with 128 of 256 experts per layer in CPU DRAM (TP=1, EP disabled). Curves report length-wise phase time divided by token-wise phase time for the same placement and batch size. Random offloading selects experts randomly, whereas hot expert offloading selects those favored by length-wise replay. The dotted line denotes equality.}
    \label{fig:moe-offload}
\end{figure*}

\paragraph{Offloading amplifies sensitivity to token content.}
With EP disabled and TP=1, we place 128 of the 256 experts per layer in CPU DRAM.
GPU kernels access their weights directly through UVA.
Under random offloading, length-wise decode takes 0.90$\times$ and 0.82$\times$ the token-wise time at batches 16 and 32 (Figure~\ref{fig:moe-offload}).
Batches 1--8 show small differences in the opposite direction.
At batch~32, CPU-resident experts receive similar assignment shares, 51.50\% for length-wise and 50.19\% for token-wise replay, despite the different working-set sizes.
Assignment share alone therefore does not characterize the memory-access workload.
Offloading 128 length-wise hot experts widens the discrepancy: length-wise prefill and decode take 0.39$\times$ and 0.63$\times$ their respective token-wise times at batch~32.
Together, these experiments show that matching lengths can overestimate EP prefill cost while underestimating TP decode and offloaded execution costs.

\subsection{Can a Smaller Model Replay a Capable Model's Workload?}
\label{sec:exp-cross-model}

\paragraph{Natural execution can miss the intended serving regime.}
The SymPy runs in Section~\ref{sec:exp-greedy} frequently exit early or become trapped in unproductive cycles of tool calling and text generation.
Early exits omit the long contexts and KV-cache residence times that a long-horizon benchmark aims to exercise.
These non-progressing cycles instead overrepresent a degenerate control-flow pattern.
Both are valid agent behaviors, but their mixture changes the serving workload across repetitions and confounds comparisons of scheduling or cache policies.

\paragraph{A capable-model trace provides a fixed target workload.}
We collect a completed trajectory for SWE-bench instance \texttt{sympy\_\_sympy-12419} using the DeepSeek-V4-Pro API and Mini-SWE-Agent, then replay it on Qwen3.6-35B-A3B.
Because the tokenizers differ, replay preserves messages, tool observations, and invocation dependencies, then fixes the resulting target-token sequence.
All 132 recorded turns and tool calls are retained, without requiring source and target token counts to match.

Natural execution takes $258.74\pm297.37$ seconds with a CV of 114.93\%, while cross-model replay with fixed tool delays takes $261.21\pm1.17$ seconds with a CV of 0.45\%.
The CV decreases by 257.1$\times$ using unrounded values, with a similar mean duration.
This result demonstrates repeatable execution of a completed long-horizon trace on the target, rather than autonomous task success or speedup.
Representativeness beyond this single trajectory requires a broader trace collection.

\section{Related Work}
\label{sec:related-work}

\paragraph{Nondeterminism in LLM inference.}
Greedy decoding removes sampling randomness, but it does not make model execution invariant to the serving configuration.
Changes in batch shape, hardware, or parallel reduction order can introduce numerical differences that flip an argmax decision.
Autoregressive decoding then propagates one changed token into a different continuation~\citep{yuan2026understanding,he2025nondeterminism}.
Batch-invariant kernels~\citep{he2025nondeterminism,qiang2026dash,zhang2025deterministic} remove this source of variation by enforcing a consistent computation.
These techniques improve output reproducibility, but they do not replace controlled workload replay for serving-system evaluation.
Requiring deterministic kernels constrains the kernels, parallelization strategies, and routing implementations available to an experiment.
\SYS{} instead fixes selected workload properties while each system retains its native execution path.
It does not require bitwise-identical activations or logits.

\paragraph{Agent traces and replay.}
TraceLab~\citep{zhu2026tracelab} collects real-world coding-agent sessions and characterizes their long contexts, repeated tool use, tool-latency tails, and prefix-cache behavior.
Its traces describe realistic workloads, but its goal is workload characterization rather than replaying model execution as a controlled serving experiment.
AgentRR~\citep{feng2025agentrr} records agent interactions and summarizes successful traces into reusable experiences that guide later agent behavior.
This abstraction targets reliability, reuse, and large--small model collaboration.
Because it may change subsequent execution, it does not define an identical serving workload.

Concurrent work XPerf~\citep{wang2026benchmarking} records input and output token IDs, invocation dependencies, and tool durations for agentic serving benchmarks and also replays accepted-token counts for speculative decoding.
Both XPerf and \SYS{} use recorded trajectories to prevent live agent variation from dominating serving comparisons.
Our study focuses on the additional distinction between length, token, route, and tool-delay controls, and on executing a capable model's trajectory with a compatible smaller model for long-horizon experiments.

\paragraph{MoE router replay.}
Rollout Routing Replay~\citep{ma2025r3} records inference-time routing distributions and reuses them in the training engine to reduce rollout--training mismatch during MoE reinforcement learning.
Megatron-LM~\citep{nvidia2026routerreplay} similarly supports loading and enforcing recorded expert assignments for determinism, debugging, and profiling in training and inference.
These mechanisms establish that expert selections can be recorded and enforced, but they target training consistency or isolated profiling rather than comparative serving benchmarks.

\section{Discussion and Conclusion}
\label{sec:conclusion}

\paragraph{Scope.}
Fixed-trace comparisons measure the cost of executing recorded behavior.
They complement live evaluation when optimizations change output quality, agent decisions, or tool-resource demand.
For example, replaying a recorded response cannot establish that quantization or approximate KV compression preserves task quality.
Speculative decoding requires separate treatment of proposals, verification, and acceptance, which our autoregressive evaluation does not cover.
Since replay fixes trajectories during measurement, its coverage still depends on collecting diverse, representative traces.

\paragraph{Conclusion.}
Agent serving benchmarks need to distinguish execution improvements from changes in the work being executed.
\SYS{} controls token histories, optional logical MoE assignments, and tool delays while preserving model execution and system scheduling freedom.
Our experiments show that greedy decoding changes the realized workload and that random-token replay distorts MoE expert access patterns and execution costs.
A capable-model trace also provides a repeatable long-horizon workload on a smaller target model.
These results support token-wise replay as a basis for controlled serving comparisons.

\label{page:main-end}
\clearpage
\section*{AI Use Statement}
The authors used generative AI tools to edit the manuscript, review literature, develop the experimental protocol, and write figure-generation code.
The authors are responsible for the accuracy of all claims, citations, and results.

\bibliography{references,references_agentreplay}
\bibliographystyle{vendor/iclr2027/iclr2027_conference}

\end{document}